\documentclass[aps,prl,reprint,groupedaddress]{revtex4-1}
\usepackage{amsmath}
\usepackage{natbib}
\usepackage{graphicx, subfigure}
\usepackage{hyperref}  %% links

\begin{document}

%\preprint{}

%\title{Negative contribution from turbulence to magnetic diffusivity in a liquid sodium experiment}
\title{Turbulence Reduces Magnetic Diffusivity in a Liquid Sodium Experiment}


%\thanks{}
\author{Simon Cabanes}
%\thanks{}
\author{Nathana\"el Schaeffer}
\author{Henri-Claude Nataf}
\email[]{henri-claude.nataf@ujf-grenoble.fr}
\affiliation{Univ. Grenoble Alpes, ISTerre, F-38000 Grenoble, France}
\affiliation{CNRS, ISTerre, F-38000 Grenoble, France}
%\affiliation{IRD, ISTerre, F-38000 Grenoble, France}


\date{\today ~-- published in \textit{Phys. Rev. Lett.} \textbf{113}, 184501 (2014)}

\begin{abstract}
The contribution of small scale turbulent fluctuations to the induction of mean magnetic field is investigated in our liquid sodium 
spherical Couette experiment with an imposed magnetic field.
An inversion technique is applied to a large number of measurements at $Rm \approx 100$ to obtain radial profiles of the  $\alpha$ and $\beta$ effects and maps of the mean flow.
It appears that the small scale turbulent fluctuations can be modeled as a strong contribution to the magnetic diffusivity that is negative in the interior region and positive close to the outer shell.
Direct numerical simulations of our experiment support these results.
The lowering of the effective magnetic diffusivity by small scale fluctuations implies that turbulence can actually help to achieve self-generation of large scale magnetic fields.

\end{abstract}

%\keywords{}

\maketitle

%\section{Introduction (NS)}

The Earth, the Sun and many other astrophysical bodies produce their own magnetic field by dynamo action, where the induction of a magnetic field by fluid motion overcomes the Joule dissipation.
In all astrophysical bodies, the conducting fluid undergoes turbulent motions, which
can also significantly affect the induction of a large-scale magnetic field by either enhancing it or weakening it.
It is therefore of primary interest to quantify the role of these fluctuations in the dynamo problem.

The induction equation for the mean magnetic field $\langle \mathbf{B} \rangle$ reads:
\begin{equation}\label{eq:F_induction}
\frac{\partial \langle \mathbf{B}\rangle}{\partial t} = \mathbf{\nabla} \times ( \langle \mathbf{U}\rangle \times \langle \mathbf{B}\rangle + \mathcal{E} ) + \eta \Delta \langle \mathbf{B}\rangle 
\end{equation}
where $\langle U \rangle$ is the mean velocity field, $\eta = (\mu_0 \sigma)^{-1}$ is the magnetic diffusivity (involving the magnetic permeability $\mu_0$ and the conductivity of the fluid $\sigma$), and $\mathcal{E} = \langle \mathbf{\tilde u} \times \mathbf{\tilde b} \rangle$ is the mean electromotive force (emf) due to small scale fluctuating magnetic $\mathbf{\tilde b}$ and velocity $\mathbf{\tilde u}$ fields.
The relative strength between the inductive and dissipative effects is given by the magnetic Reynolds number $Rm = UL/\eta$ ($U$ and $L$ are characteristic velocity and the characteristic length-scale).
When there is a scale separation between the turbulent fluctuations and the mean flow, we can follow the mean-field theory
and expand the emf in terms of mean magnetic quantities: $\mathcal{E} = \alpha \langle\mathbf{B}\rangle  - \beta \nabla \times \langle\mathbf{B}\rangle$.
For homogeneous isotropic turbulence, $\alpha$ and $\beta$ are scalar quantities.
%This is of particular interest when a large shear flow induces a toroidal magnetic field, known as an $\omega$-effect. It results from non-axisymmetric $\alpha$ induction that it can generate the needed toroidal currents to reinforce the initial poloidal field and lead to dynamo action.
$\alpha$ is related to the flow helicity and results in an electrical current aligned with the mean magnetic field, whereas $\beta$ can be interpreted as a turbulent diffusivity effectively increasing ($\beta > 0$) or decreasing ($\beta < 0$) electrical currents.
The effective magnetic diffusivity $\eta_{eff} = \eta + \beta$ can have tremendous effects on energy dissipation and on dynamo action by reducing or increasing the effective magnetic Reynolds number $Rm_{eff} = UL/\eta_{eff}$.

However, direct determination of these small-scale contributions remains a challenging issue for experimental studies and numerical simulations.\\

The first generation of dynamo experiments were designed to show that turbulent flows with strong geometrically-imposed helicity could self-generate their own magnetic fields.
Since the success of Riga \citep{gailitis01} and Karlsruhe \citep{stieglitz01} dynamos, several other liquid metal experiments have sought to overcome the effects of magnetohydrodynamic turbulence in less constrained, more geophysically relevant  flow geometries.
Unfortunately, dynamo action remains elusive, and the effective contribution of small-scale motions to large-scale magnetic fields remains poorly understood, though the small-scale motions seem to work against dynamo action \cite{spence06,frick10}.

%\citet{spence06} were the first to document a negative global response of the $\alpha$-effect on the induced magnetic dipole in the Madison experiment.
In the Perm torus-shaped liquid sodium experiment, the effective magnetic diffusivity was inferred from phase shift measurements of an alternating magnetic signal, indicating turbulent increases in magnetic diffusivity of up to $\approx 30 \%$ \cite{frick10}.
%More recently in the Madison facility, \citet{rahbarnia12} reported direct measurements of the local emf, which account for both $\alpha$ and $\beta$ contributions. They observed that turbulent contributions were dominated by the $\beta$-effect which was responsible of a magnetic diffusivity increase of about $\approx 30 \%$.
The Madison experiment, a sphere containing two counter-rotating helical vortices, found that an externally applied magnetic field was weakened by about $20 \%$ at $Rm = 130$, which they interpreted as a negative global $\alpha$-effect \cite{spence06}.
The installation of an equatorial baffle was found to reduce the amplitude of the largest-scale turbulent eddies and hence the $\alpha$-effect \cite{kaplan11}.
In the same set-up, \citet{rahbarnia12} measured the local emf directly, finding contributions from both $\alpha$ and $\beta$, but with a dominant $\beta$-effect.
They reported an increase in magnetic diffusivity of about $30\%$.
The Von Karman Sodium  experiment, a cylinder containing another two-vortex liquid sodium flow, reported a magnetic diffusivity increase of about $100 \%$ \cite{ravelet12}.
%\citet{nataf13} pointed out that very similar results are obtained in different $Rm$ regime. It is indeed likely that experimental device geometry bring additional constraint that we do not overcome. \\

%\section{Set-up, data and inversion  (HCN)}

\begin{figure}
\includegraphics[width=\linewidth]{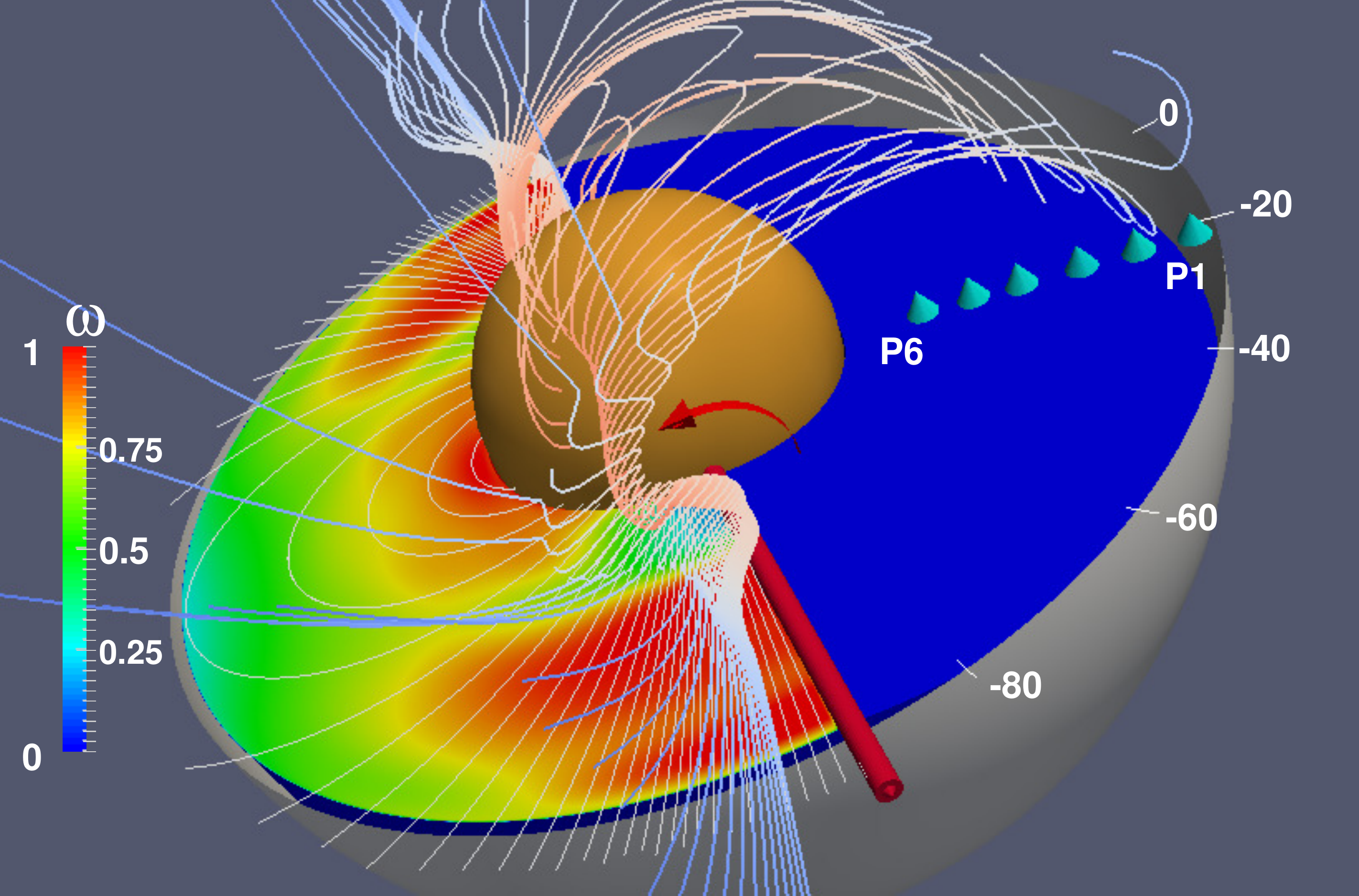}
\caption{Sketch of the DTS experiment with its liquid sodium contained between an outer stainless steel shell (grey, with latitude labels in degrees) and an inner copper sphere (orange), which spins as indicated by the red arrow around the vertical rotation axis (here tilted for clarity). \textit{left half of the sphere}: the field lines of the dipolar magnetic field imposed by the central magnet are drawn on top of the contour map of the fluid angular velocity $\omega$ (normalized by that of the inner sphere) inverted from data measured for $Rm=94$. \textit{right half of the sphere}: field lines of the total reconstructed magnetic field. The field lines are strongly distorted by the flow ($\omega$-effect). The blue cones mark the radial positions of the 6 magnetometers $P1$ ($r=radius/r_o=0.99$) to $P6$ ($r=0.50$), which measure the azimuthal magnetic field. They can be placed at 4 different latitudes (here $-20^\circ$).
\label{fig:U and B map}}
\end{figure}

We analyze data from the Derviche Tourneur Sodium experiment (DTS), a magnetized spherical Couette flow experiment  sketched in Figure \ref{fig:U and B map}.
Forty liters of liquid sodium are enclosed between an inner sphere (radius $r_i=74$mm) and a concentric outer stainless steel shell (inner radius $r_o=210$mm).
The inner sphere can rotate around the vertical axis at rates up to $f=30$Hz, yielding a maximal value of $94$ for the magnetic Reynolds number defined as $Rm=2 \pi f r_o^2/\eta$.
The inner sphere consists of a copper shell containing a strong permanent magnet, which produces a, mostly dipolar, magnetic field pointing upwards along the rotation axis.
The intensity of the magnetic field decreases from $B_i \simeq 180$mT at the equator of the inner sphere to $B_o \simeq 7.1$mT at the equator of the outer shell.
More details are given in \cite{brito11}.

In a recent study \citep{cabanes14}, we developed a new strategy to determine the mean velocity and induced magnetic fields.
Following earlier works \cite{brito11,nataf13}, we collect ultrasound Doppler velocity profiles, electric potential measurements, global torque data, and measurements of the induced magnetic field inside the sodium layer, to reconstruct meridional maps of the mean flow and magnetic field at a given $Rm$, taking into account the link established by the induction equation.
But we further constrain these fields by analyzing the response of the fluid shell to a time-periodic magnetic field, as in \citet{frick10}.
In our case, the time-periodic signal simply results from the rotation of our central magnet, whose small deviations from axisymmetry produce a field varying at the rotation frequency and its harmonics.
We have expanded the complete magnetic potential of the magnet in spherical harmonics up to degree 11 and order 6, which we then use to compute the solution of the time-dependent induction equation.
The predictions for a given mean velocity field are compared to actual magnetic measurements inside the sodium shell at 4 latitudes and at 6 radii, as depicted in Figure \ref{fig:U and B map}.
%Extending the numerical spherical spectral code developed by \citet{figueroa13},
We construct a non-linear inversion scheme of the induction equation to retrieve the mean axisymmetric (and equatorially-symmetric) toroidal and poloidal velocity fields that minimize the difference between the predictions and all measurements at a given rotation rate $f$ of the inner sphere.
 \citet{cabanes14} discuss in detail the solutions and fits for $Rm = 28$.

In the present study, we extend the analysis to the largest available $Rm=47$,  $72$ and $94$ (see Table \ref{tab:parametres} for details).
Figure \ref{fig:U and B map} displays a meridional map of the angular velocity inverted for $Rm = 94$, and the field lines of the predicted magnetic field.
%They confirm that the azimuthal flow obeys Ferraro law \cite{ferraro37} near the equator of the inner sphere, where the magnetic field is strong.
They confirm that, near the equator of the inner sphere where the magnetic field is strong, the angular velocity stays nearly constant along magnetic field lines (Ferraro law \cite{ferraro37}).
That region displays super-rotation, while the flow becomes more geostrophic further away from the inner sphere.

 \begin{table}%[H] add [H] placement to break table across pages
 \caption{For each inner sphere rotation rate $f$, we list the corresponding $Rm$, the total number $Np$ of free parameters we invert for,  the total number $Nd$ of data points including mean measurements and time-varying magnetic data, and the associated global normalized misfit $\chi$ (the error-weighted rms difference between observations and predictions). The number of data points is much smaller at high $Rm$ as ultrasound Doppler velocimetry is not operational. Values in brackets are the numbers obtained when we do not invert for $\alpha$ and $\beta$.
 \label{tab:parametres}}
 \begin{ruledtabular}
\begin{tabular}{ccccccccc}
$f$ (Hz) & $Rm$ & $Np$ & $Nd$ & $\chi$  \\
\colrule
$-9$ & $28$ & $108 \, (96)$ & $1130$ & $1.5 \, (1.8)$ \\
$-15$ & $47$ & $108 \, (96)$ & $440$ & $2.5 \, (3.3)$ \\
$-23$ & $72$ & $60 \, (48)$ & $230$ & $2.5 \, (4.9)$ \\
$-30$ & $94$ & $60 \, (48)$ & $230$ & $2.9 \, (5.9)$ \\
\end{tabular}
 \end{ruledtabular}
 \end{table}
 
However, the mean velocity field alone does not fully account for the measured mean magnetic field.
\citet{figueroa13} point out that velocity fluctuations invade the interior of the shell in DTS as the rotation rate $f$ increases, and that magnetic fluctuations always get larger towards the inner sphere because of the strong imposed magnetic field there.
We therefore extend our previous approach \citep{cabanes14} to take into account the contribution of turbulent fluctuations to the mean magnetic field.
Following earlier attempts \citep{spence06, frick10, rahbarnia12}, we choose to invert for $\alpha$ and $\beta$, but since we expect that fluctuations will strongly depend upon the intensity of the mean magnetic field, we allow them to vary with radius.
Note that time-varying magnetic signals are particularly sensitive to the effective magnetic diffusivity, hence to $\beta$ \citep{frick10,tobias13}.

We thus simultaneously invert for the mean axisymmetric toroidal velocity field $U_T(r,\theta)$ and for radial profiles $\alpha(r)$ and $\beta(r)$.
$U_T$ is decomposed in spherical harmonics up to $l_{max}=8$ (m=0) and in Chebychev polynomials in radius up to $n_{max}=11$.
$\alpha(r)$ and $\beta(r)$ are projected on Chebychev polynomials up to $k_{max}=5$, leading to:
\begin{equation}\label{eq:emf description}
 \mathcal{E}(r) = \sum_{k=0}^{5} T_k(r)  \left( \alpha_k \langle\mathbf{B}\rangle - \beta_k \mathbf{\nabla} \times \langle\mathbf{B}\rangle \right),
\end{equation}
where $T_k$ is the degree $k$ Chebychev polynomial of the first kind and $\langle\mathbf{B}\rangle$ is the total mean magnetic field, solution of equation \eqref{eq:F_induction}.
Since the inversion is slightly non-linear, we use the linearized least-square Bayesian method of \citet{tarantola82}, taking the {\it a posteriori} velocity model from a lower $Rm$, upscaled to the new $Rm$, as the \textit{a priori} velocity model.
We choose a zero value as the {\it a priori} model for all $\alpha_k$ and $\beta_k$.
The poloidal velocity field is at least one order of magnitude smaller than the toroidal one.
We do not invert for it at $Rm = 72$ and $94$ but we include in the direct model a meridional flow up-scaled from the solution obtained at $Rm=47$ \citep{cabanes14}.
We find that solving for the emf, which adds only $12$ degrees of freedom, reduces the global normalized misfit significantly (see Table \ref{tab:parametres}).

%\section{experimental results  (HCN)}

\begin{figure} %--------------------------------- figures des effets alpha & beta ------------------------------%
     \begin{center}
        \subfigure{%[Caption of First Figure]{%
        %    \label{fig:alpha effect}
            \includegraphics[width=\linewidth]{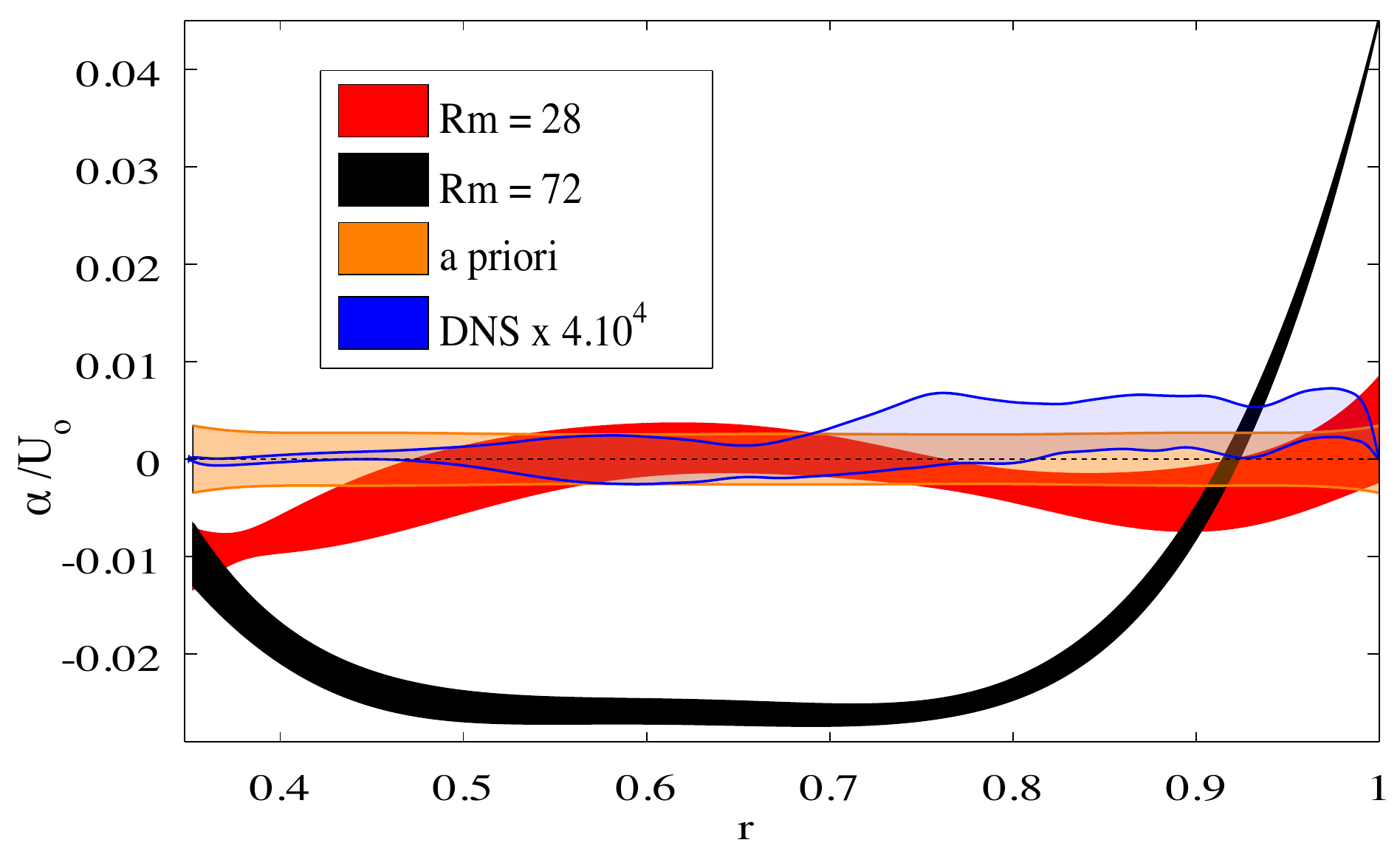}
        }\\%
        \subfigure{%[caption]{%
        %   \label{fig:beta effect}
           \includegraphics[width=\linewidth]{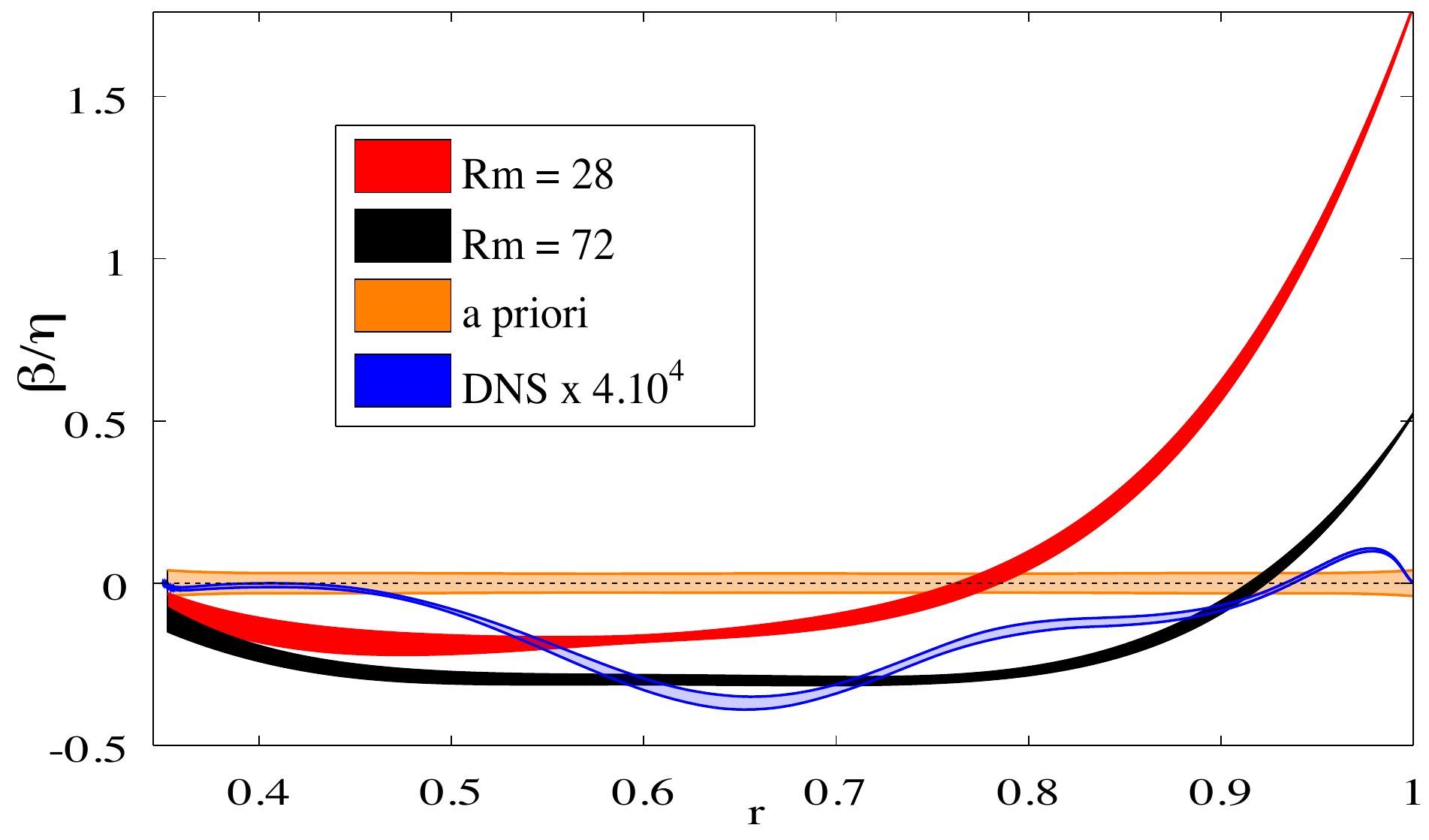}
           }%
    \end{center}
    \caption{%
Radial profiles of the $\alpha$-effect (a) and $\beta$-effect (b) with their error bars, obtained by the inversion of DTS data for two magnetic Reynolds number: $Rm=28$ and $72$. The \textit{a priori} null profile, along with its error bar, is also drawn. The blue curve shows the $\alpha(r)$ and $\beta(r)$ profiles retrieved from a numerical simulation of the DTS experiment at $Rm=29$ and $Re=2.9 \times 10^4$, blown up by a factor $4 \times 10^4$.
     }%
   \label{fig:alpha and beta effect}
\end{figure}%------------------------------------------- figures des effets alpha & beta ------------------------------%

Figure  \ref{fig:alpha and beta effect} shows the radial profiles of $\alpha$ and $\beta$ (with their \textit{a posteriori} model errors) produced by the inversion of data at $Rm=28$ and $72$.
The profiles for $Rm = 94$ (not shown) are almost the same as for $Rm=72$.
$\alpha$ is normalized by $U_0=2 \pi f r_o$, and $\beta$ by $\eta$.
For the lower $Rm$ value, we observe practically no $\alpha$-effect, while the $\beta(r)$ profile indicates that the $\beta$-effect increases strongly when going from the Lorentz-force-dominated inner region to the Coriolis-force-dominated outer region.
It reaches values of $1.7\eta$ near the outer boundary, where velocity fluctuations are strongest \citep{figueroa13}.
For the higher $Rm$, some $\alpha$-effect is required to match the data over most of the fluid domain.
The $\beta(r)$ profile displays strongly negative values (down to $-0.3\eta$) over almost the complete fluid shell, but rises sharply to positive values near the outer boundary.
%\citet{figueroa13} point out that velocity fluctuations invade the interior of the shell in DTS as the rotation rate $f$ increases, and that magnetic fluctuations always get larger towards the inner sphere because of the strong imposed magnetic field there.

 \begin{figure}
\includegraphics[width=\linewidth]{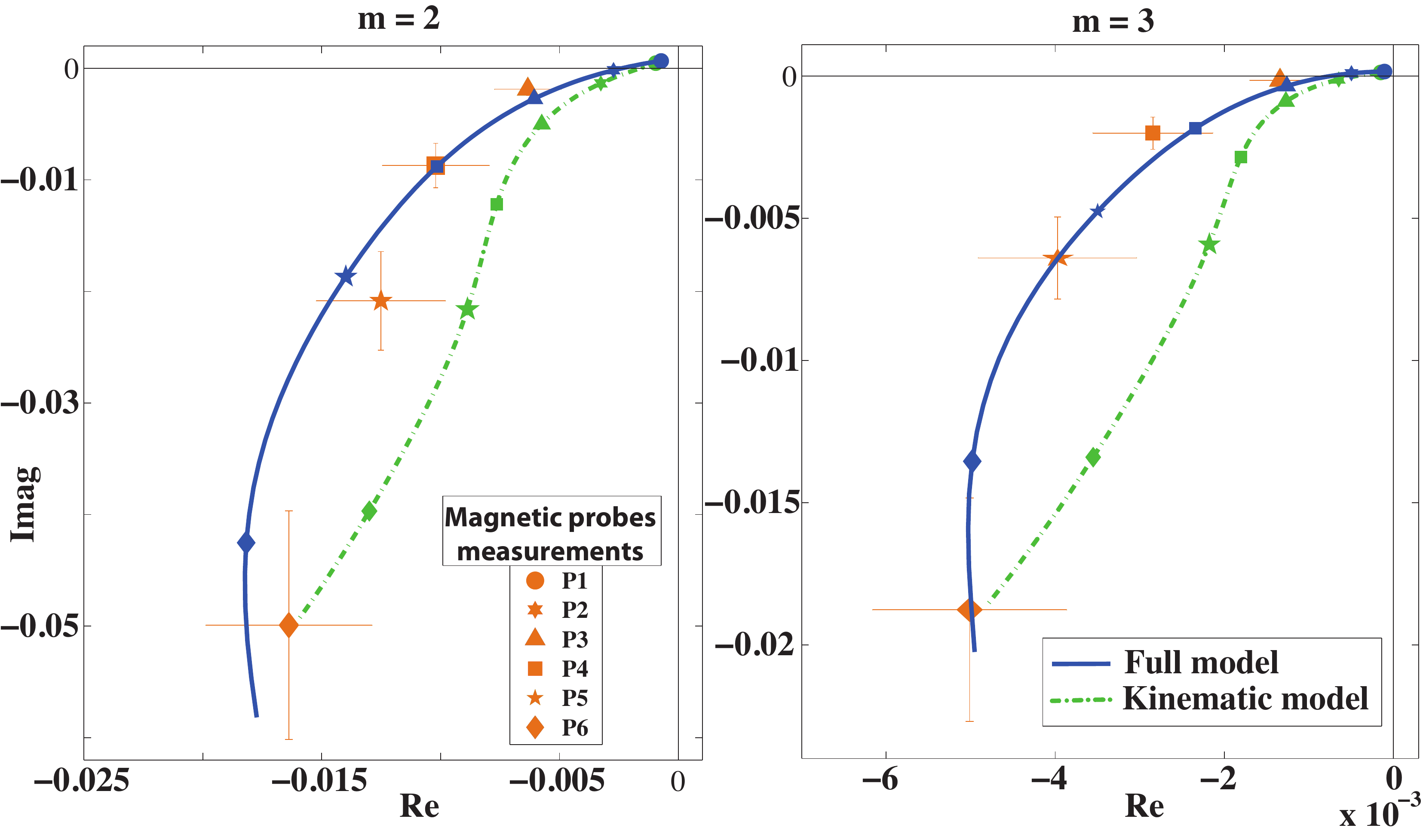}
\caption{Measurements and model fits for an example of time-varying magnetic signals measured at $2f$ (m=2) and $3f$ (m=3) frequencies, for a rotation rate of the inner sphere $f=-23$Hz ($Rm = 72$).
See text for explanations.
\label{fig:azimuthal modes}}
\end{figure}

\begin{figure*}
\includegraphics[width=0.75\linewidth]{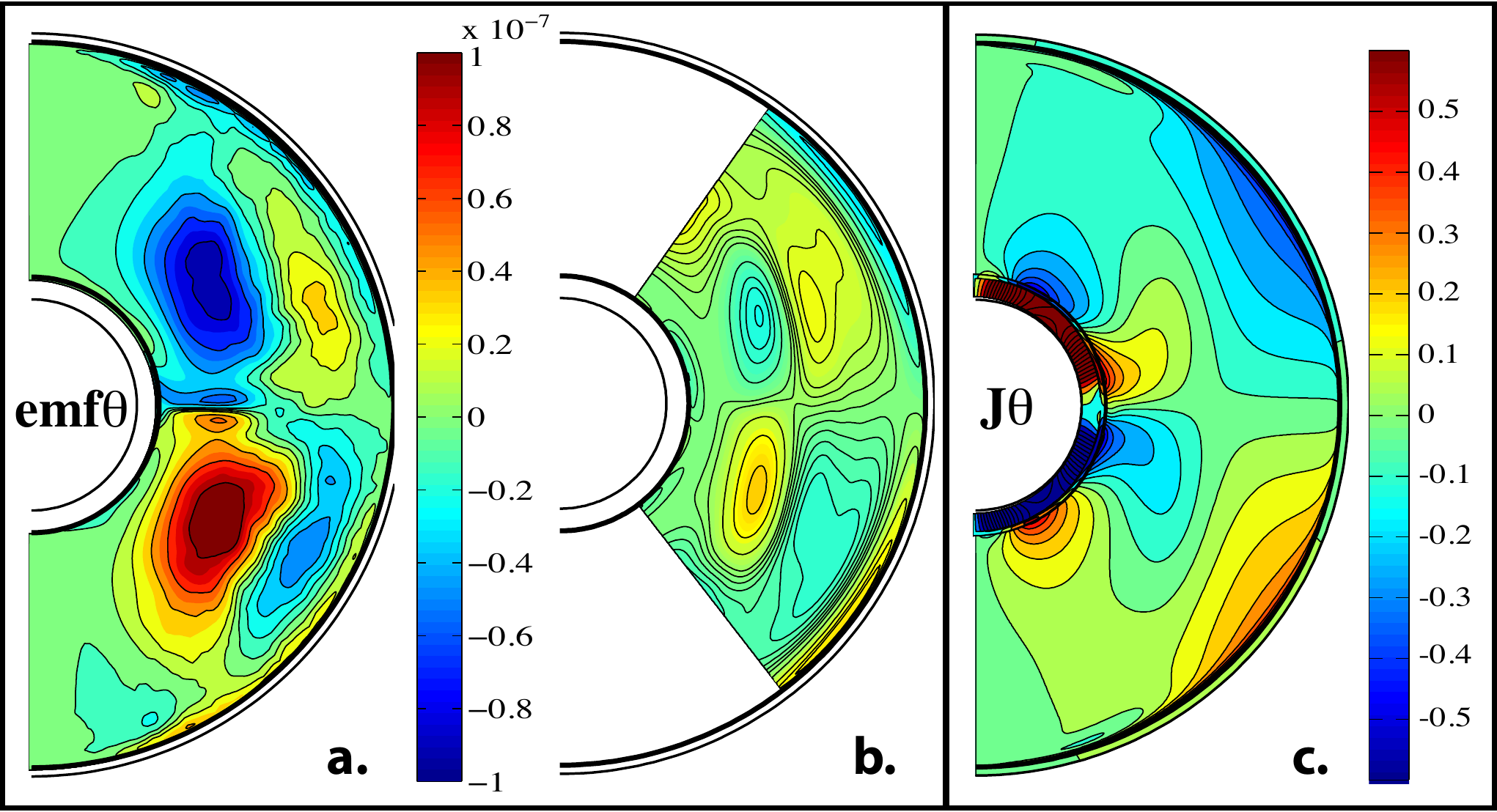}
\caption{Meridional cross section contour maps showing orthoradial component ($\theta$) of emf $\mathcal{E}$ and of electrical current $\langle\mathbf{J}\rangle$.
(a) Averaged emf $\mathcal{E}_t$ obtained from DNS.
(b) Reconstructed emf $\mathcal{E}_{\alpha\beta}$ from inverted $\alpha$ and $\beta$ profiles. High latitudes (white area) are excluded from the least-square fit. % to be coherent with the determination of $\alpha$ and $\beta$ in the experiment.
(c) Mean electrical current from DNS.
%All quantities are dimensionless \textbf{explain how dimensionless}.
\label{fig:emf}}
\end{figure*}

The introduction of the $\alpha$- and $\beta$-effects clearly improves the fit to the measurements.
We illustrate this in Figure \ref{fig:azimuthal modes}, which compares the prediction of our model, with and without the $\alpha$ and $\beta$ terms, to the measurements of the time-varying signals for $f=-23$Hz ($Rm=72$), at a given latitude ($-20^\circ$).
There, a sleeve intrudes into the sodium volume and records the azimuthal component of the magnetic field at 6 different radii labeled P1 to P6 (as drawn in Figure \ref{fig:U and B map}).
When the inner sphere spins, small deviations of its magnetic field from axisymmetry produce a magnetic signal that oscillates at the rotation frequency $f$ and its overtones.
Here we focus on the $2f$ and $3f$ overtones caused by the $m=2$ and $m=3$ heterogeneities of the magnet.
We measure the phase and amplitude of the time-varying magnetic signals at all 6 radii and plot them (with their error bars) in the complex plane, normalized by $B_0$ (the intensity of the imposed magnetic field at the equator of the outer shell).
When the inner sphere is at rest, we record only the magnet's potential field weakening with increasing distance.
Advection and diffusion completely distort this pattern when the inner sphere spins.
The blue solid line displays the prediction from our full model of these magnetic signals from the largest values at the inner sphere boundary ($r=r_i$) to small values at the outer sphere ($r=r_o$).
Symbols mark the radial positions of the P6 to P1 magnetometers.
The green dashed line is the trajectory predicted by our model when we remove the $\alpha$ and $\beta$ terms.
This altered model fails to produce the observations, indicating that the $\beta$-effect that we retrieve contributes significantly to the measured signals.

In addition to the inversion of experimental measurements, we perform direct numerical simulations (DNS) of the experiment. Our code, based on spherical harmonic expansion \cite{schaeffer13} and finite differences in radius, has already been used to simulate the experiment. We restarted the most turbulent computation of \citet{figueroa13} with a new imposed magnetic field containing the additional non-axisymmetric and non-dipolar terms.
This simulation reaches $Re=2\pi f r_o^2/\nu = 2.9 \times 10^4$ ($\nu$ is the kinematic viscosity), $Rm=29$ and a magnetostrophic regime close to that of the experiment \cite{brito11}.
Turbulence is generated by the destabilization of the outer boundary layer, yielding plumes that penetrate inward to regions of stronger magnetic fields. There, the velocity fluctuations are damped, but the associated magnetic fluctuations are stronger \cite{figueroa13}.
Six snapshots of the fields are saved every five turns.
After we have reached a statistically steady regime, we average the fields over 162 turns of the inner sphere to obtain $\langle \mathbf{B}\rangle$ and $\langle \mathbf{U} \rangle$.
It is then straightforward to compute the mean emf $\mathcal{E} = \langle \mathbf{\tilde u} \times \mathbf{\tilde b} \rangle$ where fluctuating fields are obtained from the difference between a snapshot and the time- and longitude-averaged field.

Meridional maps of the mean emf $\mathcal{E}_t$ are obtained and the latitudinal component is displayed in Figure \ref{fig:emf}a.
The $\alpha$ and $\beta$ profiles that best explain this mean emf (least-square solution of equation \ref{eq:emf description} excluding high latitudes) are shown in Figure \ref{fig:alpha and beta effect}.
We estimate the error bar on the profiles as the standard deviation of emfs computed from 5 subsamples of 40 snapshots.
One component of the emf $\mathcal{E}_{\alpha\beta}$ computed with these $\alpha$ and $\beta$ profiles is shown in Fig. \ref{fig:emf}b, and can be compared to the actual emf $\mathcal{E}_t$ (Fig. \ref{fig:emf}a).
Although the $\alpha$ and $\beta$ profiles do not explain all of the mean emf, most features are recovered.
Other components exhibit a similar behavior (not shown).

The parity (symmetry with respect to the equatorial plane) of the emf and of $\langle\mathbf{J}\rangle$ are clearly even (Fig. \ref{fig:emf}c), while $\langle\mathbf{B}\rangle$ is odd.
This is in line with the fact that the DNS, just like the experiments at the lowest $Rm$, predicts no $\alpha$-effect (see Fig. \ref{fig:alpha and beta effect}a).
This might seem surprising given that the mean flow displays helicity.
However, if we split the velocity fluctuations into even ($\mathbf{\tilde u^+}$) and odd ($\mathbf{\tilde u^-}$) parity, we see that their interaction with the mean odd magnetic field generates odd ($\mathbf{\tilde b^-}$) and even ($\mathbf{\tilde b^+}$) magnetic fluctuations, respectively.
The resulting emf $\mathcal{E} = \mathbf{\tilde u} \times \mathbf{\tilde b}$ is therefore always even, if the odd and even velocity fluctuations are uncorrelated. This is likely true in the low $Rm$ regime.
The fact that the higher $Rm$-experiments require a non-zero $\alpha$-effect (Fig. \ref{fig:alpha and beta effect}a) reveals that the velocity fluctuations are interacting with an already-distorted larger-scale magnetic field, or that correlations between the two parities become non-zero.

The dipolar component of the induced magnetic field predicted by our full model is small but non-zero at the surface of the outer shell, even when the $\alpha$-effect is negligible.
\citet{spence06} have shown that an axisymmetric flow interacting with an axisymmetric magnetic field cannot produce an external dipole.
This remains true if fluctuations only result in a homogeneous $\beta$-effect.
Even with a radially-varying $\beta$-effect as we obtain here, an external dipole can be produced only if a meridional flow is present.

The most striking feature of the $\beta(r)$ profiles we retrieve is the strong negative values (down to $-0.3\eta$) that span a large portion of the liquid sodium shell, especially at large $Rm$ (see Fig. \ref{fig:alpha and beta effect}).
The DNS supports this result, showing that it is not an artifact of considering only a radial dependence for $\alpha$ and $\beta$.
The much lower amplitude of $\beta$ in the DNS is due to a Reynolds number 300 times smaller than that in the experiment, suggesting that $\beta$ may scale with $Re^2$ (but see the Erratum below).
Although negative $\beta$ values, and hence reduced magnetic diffusivity, are not unexpected \citep{Zheligovsky03,brandenburg08, giesecke14, lanotte1999}, it is the first time that they are observed in experiment.
Our DTS experiment combines a strong imposed magnetic field and strong rotation.
These could be the ingredients that lead to this behavior.
Were $\beta$ to become even more negative, it might promote dynamo action.\\

\section*{Erratum}

In our original letter \citep{Cabanes14b}, there was an inconsistency in the sign convention used for $\beta$.
The typos have been corrected in the present document and did not affect the profiles inverted from our experimental data.
Unfortunately the wrong sign for $\beta$ was used when analyzing the results of the numerical simulations.
In addition, a mistake in the normalization of the EMF computed from the simulations makes it appear 561 times smaller than it actually is.
The much lower amplitude of $\beta$ in the DNS was interpreted as a suggestion for $\beta$ scaling as $Re^2$ (the square of the Reynolds number).
Instead, the correct amplitude is actually in line with a $\beta$ effect increasing proportionally to the Reynolds number: $\beta \sim Re$.

\begin{figure}
     \begin{center}
      \vspace{1cm}
      \includegraphics[width=\linewidth]{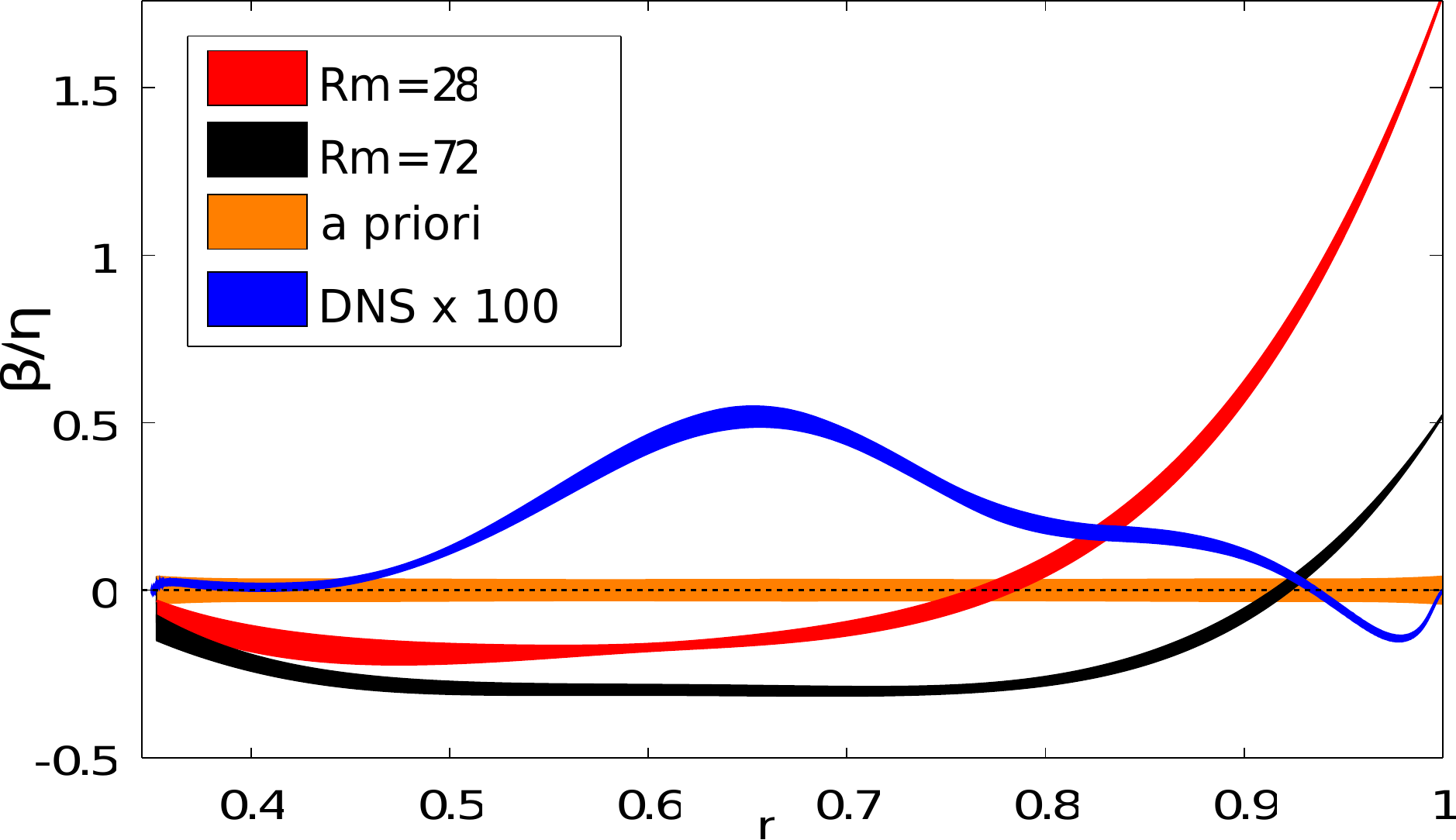}
    \end{center}
    \caption{(color online)
Radial profiles of the $\beta$-effect with their error bars (the line thickness), obtained by the inversion of DTS data for two magnetic Reynolds number: $Rm=28$ and $72$.
The \textit{a priori} null profile, along with its error bar, is also drawn.
$\beta$ is normalized by the molecular magnetic diffusivity $\eta$.
The blue curve shows the $\beta(r)$ profile retrieved from a numerical simulation of the DTS experiment at $Rm=29$ and $Re=2.9 \times 10^4$, blown up by a factor $100$.
     }
   \label{fig2b_correct}
\end{figure}

Figure \ref{fig2b_correct} replaces the original Fig. \ref{fig:alpha and beta effect}b found in our letter.
After making the corrections, the numerical simulations are no more in good agreement with the $\beta$-effect found in the experiment, as they now have more or less opposite signs.
%The $\beta$ profile that best explains the induced magnetic field in the numerical simulations still exhibit a small region of negative magnetic diffusivity, but it is now located near the outer boundary.

%On the bright side, since the publication of our letter, we have been pointed to a theoretical study that demonstrates the existence of incompressible flow producing negative $\beta$-effect \cite{lanotte1999}. They give the concrete example of a Taylor-Green type flow which involves large-scale shear as in our experiment.

We acknowledge that our numerical simulations, performed at much lower Reynolds number, do not show the same behavior as the experimental data.
These data remain however best explained by a reduced effective magnetic diffusivity due to turbulent fluctuations.

\section*{Acknowledgments}

This work was supported by the National Program of Planetology of CNRS-INSU under contract number AO2013-799128, and by the University of Grenoble.
Most computations were performed on the Froggy platform of CIMENT (\url{https://ciment.ujf-grenoble.fr}), supported by the Rh\^one-Alpes region (CPER07\_13 CIRA), OSUG@2020 LabEx (ANR10 LABX56) and Equip@Meso (ANR10 EQPX-29-01). We thank two anonymous referees and Elliot Kaplan for useful suggestions.

%We are grateful to Johann Herault and Frank Stefani for pointing out a misprint in the version of this paper published in PRL (wrong sign appearing in front of $\beta$ in two equations).

We wish to thank Johann H\'erault and Frank Stefani for pointing out the inconsistency of the sign of $\beta$ in our original letter.
We also thank Elliot Kaplan for spotting the normalization issue in the numerical simulation post-processing code.

\bibliography{biblio_simon_300714}

\end{document}